# Title: Superconductivity in freestanding infinite-layer nickelate membranes

**Authors**

Shengjun Yan[1,2,†], Wei Mao[1,2,†], Wenjie Sun[1,2,†], Yueying Li[1,2], Haoying Sun[1,2], Jiangfeng Yang[1,2], Bo Hao[1,2], Wei Guo[1,2], Leyan Nian[3], Zhengbin Gu[1,2], Peng Wang[4] and Yuefeng Nie[1,2,*]

**Affiliations**

[1] National Laboratory of Solid State Microstructures, Jiangsu Key Laboratory of Artificial Functional Materials, College of Engineering and Applied Sciences, Nanjing University, Nanjing, 210093, P. R. China.

[2] Collaborative Innovation Center of Advanced Microstructures, Nanjing University, Nanjing, 210093, P. R. China.

[3] Suzhou Laboratory, Suzhou, 215125, P. R. China.

[4] Department of Physics, University of Warwick, Coventry CV4 7AL, UK.

† These authors contributed equally to this work.

* Corresponding author: ynie@nju.edu.cn

**Abstract**

The observation of superconductivity in infinite-layer nickelates has attracted significant attention due to its potential as a new platform for exploring high-$T_c$ superconductivity. However, thus far, superconductivity has only been observed in epitaxial thin films, which limits the manipulation capabilities and modulation methods compared to two-dimensional exfoliated materials. Given the exceptionally giant strain tunability and stacking capability of freestanding membranes, separating superconducting nickelates from the as-grown substrate is a novel way to engineer the superconductivity and uncover the underlying physics. Herein, we report the synthesis of the superconducting freestanding La$_{0.8}$Sr$_{0.2}$NiO$_2$ membranes ($T_c^{zero}$=10.9 K), emphasizing the crucial roles of the interface engineering in the precursor phase film growth and the quick transfer process in achieving superconductivity. Our work offers a new versatile platform for investigating the superconductivity in nickelates, such as the pairing symmetry via constructing Josephson tunneling junctions and higher $T_c$ values via high-pressure experiments.

**Introduction**

The discovery of superconductivity in infinite-layer nickelate films (*1*) provides a new platform for investigating the underlying mechanism of unconventional high-$T_c$ superconductivity, as they exhibit both similarities and differences compared with high-$T_c$ cuprates (*2*). The aforementioned phenomena encompass the intertwined collective excitations (*3-5*), linear-resistivity behaviors [(6)] and magnetism in ground states (*7, 8*). However, unlike cuprate superconductors first discovered in polycrystalline form (*9*), only epitaxial thin films have shown superconducting with relatively low $T_c$ values in infinite-layer nickelates (typically < 20 K) (*10-15*), which need to be further improved for physical research and practical applications. Excitingly, the recent report indicates that the $T_c^{onset}$ can be greatly enhanced in Pr$_{1-x}$Sr$_x$NiO$_2$ films by applying hydrostatic pressure, from 17 K at 0 GPa to 31 K at 12.1 GPa, without any sign of saturation before film cracking occurs (*16*). It is therefore intriguing to investigate further pressure/strain modulations in infinite-layer nickelates to determine whether $T_c$ can also surpass liquid nitrogen temperature under high pressure, similar to what has been observed in Ruddlesden-Popper nickelates La$_3$Ni$_2$O$_7$ (*17*). Additionally, despite some indirect experimental evidence suggesting a possible *d*-wave pairing form (*18-20*), the exploration of the pairing mechanism in nickelates remains primarily within theoretical predictions (*21-25*). Due to the absence of superconductivity in bulk samples (*26-28*), fabricating specific devices based on Josephson effects (*29-32*) seems impossible using available epitaxial thin films restricted by the substrates. Consequently, new forms of nickelates with more degrees of freedom are highly desired for engineering and investigating the superconductivity in nickelates.

Urgent research demands for new forms of nickelate superconductors remind us of the recent rapid development of freestanding membranes, which are synthesized by etching of the water-soluble sacrificial layer (*33*). This new epitaxial lift-off technique offers a unique opportunity to exfoliate three-dimensional (3D) perovskite materials with stronger interlayer interaction (*34-36*). The freestanding membranes provide a broad platform to construct device architectures and flexible electronics (*37*) by stacking and assembling themselves together, exhibiting great potential for exploring pairing symmetry (*30*) through twisted membranes (*38*). Furthermore, their structural flexibility and tunability make them well-suited for high-pressure experiments, thereby facilitating the achievement of higher $T_c$ values in infinite-layer nickelates. The freestanding membranes can also be mechanically attached to the external platform for applying continuous biaxial or uniaxial strains, realizing a strain state of up to 8% (*39*), far beyond than what can be attained in bulk crystals and epitaxial thin films (*40-42*). However, in order to exfoliate nickelate membranes, it is necessary to insert the water-soluble sacrificial layer that serves as an interlayer between the substrate and the target film. Given that the optimally doped precursor phase nickelates exhibit a rather extreme oxidation state of $Ni^{3.2+}$, the synthesis of perovskite phase is inherently difficult and requires specific interface engineering in order to suppress the formation of the Ruddlesden–Popper-type faults and improve the crystalline quality (*43, 44*), as these impurity phases can impede the formation of the infinite-layer structure (*45*), eliminating the superconductivity in the final reduced samples. Such critical requirement of the interface configuration exacerbates the difficulties in growing high-quality precursor phase on sacrificial layers for synthesizing freestanding membranes. Moreover, the metastable state with the infinite-layer structure can be reverted back to the precursor phase or collapse by adopting oxygens from the environment (*46*), which can be more pronounced in freestanding membranes with two exposed surfaces. For transport measurements, the film transfer process and the electrode preparation process could also induce irreversible damage to the infinite-layer structures. Therefore, while it is highly desirable, the synthesis and characterization of superconducting nickelate membranes are extremely challenging and still lacking up to date.

In this work, utilizing reactive molecular beam epitaxy (MBE), we synthesized high crystalline quality epitaxial precursor phase perovskite nickelates $La_{0.8}Sr_{0.2}NiO_3$ on the sacrificial layer $Sr_4Al_2O_7$ with an additional $SrTiO_3$ thin layer in between. By a combination of co-deposition and shuttered growth modes to engineering the interface, a fully $TiO_2$-terminated $SrTiO_3$ intermedia layer was prepared and shown to be essential for suppressing impurity phases and improving the crystalline quality of $La_{0.8}Sr_{0.2}NiO_3$. Compared to the commonly-used sacrificial layer $Sr_3Al_2O_6$, the $Sr_4Al_2O_7$ with faster dissolution rate was selected to ensure a quicker transfer process to effectively prevents oxygen backfilling in the infinite-layer membranes. By employing a well-controlled electrode transfer technique without local heating or atomic bombardment, we have observed superconductivity in the freestanding $La_{0.8}Sr_{0.2}NiO_2$ membranes with $T_c^{zero}$ of 10.9 K.

**Results**
We have successfully synthesized the high-quality precursor phase nickelates $La_{0.8}Sr_{0.2}NiO_3$ (LSNO$_3$) on the $Sr_4Al_2O_7$ (SAO$_T$) sacrificial layer by engineering the chemical composition at the interface with atomic precision. The clear reflection high-energy electron diffraction (RHEED) patterns confirm the high crystalline quality of the SAO$_T$ sacrificial layers as depicted in fig. S1, establishing a solid foundation for subsequent growth of the target compounds. The $SrTiO_3$ (STO) buffer layer with TiO$_2$-termination denoted by STO(TiO$_2$) was introduced to improve the crystalline quality of LSNO$_3$, which effectively suppresses the impurity phases reflected by diffraction spots in RHEED patterns (Fig. 1A). In comparison to the LSNO$_3$/SAO$_T$ heterostructure, where the precursor phase LSNO$_3$ was directly grown on the sacrificial layer, the STO buffer layer serves to rectify stoichiometry deviations by preventing penetration of Ni atoms into the SAO$_T$ layer which possesses a loosely arranged structure with empty sites (*33*). Furthermore, a shuttered growth method was employed to deposit [SrO] and [TiO$_2$] layers in a sequence to obtain a chemically sharp TiO$_2$-terminated surface (*47*). Then, before the growth of LSNO$_3$ films using a co-deposition method where [LaO] and [NiO$_2$] layers were deposited simultaneously, an extra [LaO] layer was grown on STO buffer layer. The insertion of an extra [LaO] layer was shown to be critical for the growth of a complete [LaO]/[TiO$_2$] configuration at the polar/non-polar (LSNO$_3$/STO) interface, which can minimize the interfacial defects and ion diffusion (*44*). As a result, the clear and bright RHEED patterns of the LSNO$_3$/STO(TiO$_2$)/SAO$_T$ heterostructure demonstrate its high surface quality, comparable to the nickelates directly grown on the STO substrate (fig. S2). In contrast, RHEED patterns of the LSNO$_3$/STO(mixed)/SAO$_T$ heterostructure, prepared by growing the STO buffer layer with mixed-termination using a co-deposition method, exhibit weak and blurred features, which is similar to the LSNO$_3$/SAO$_T$ heterostructure. The film quality is further confirmed by the X-ray diffraction (XRD) 2$\theta$–$\omega$ scans, as shown in Fig. 1B. Only the LSNO$_3$ films grown on the STO(TiO$_2$)/SAO$_T$ display the (003) diffraction peak with a minimum out-of-plane $c$ lattice constant of 3.755 Å, suggesting minimal defects in the perovskite structure (*48*). Additionally, the reciprocal space mapping (RSM) results suggest that they are fully strained to the STO substrate without any lattice relaxation (Fig. 1C). These results reveal the effective suppression of impurity phases and the improvement of precursor phase quality by interface engineering via introducing the STO(TiO$_2$) buffer layer.

The microscopic structures of LSNO$_3$ films with or without interface engineering is uncovered by using cross-sectional scanning transmission electron microscopy (STEM) in high-angle annular dark field (HAADF) mode. As the intensity of different atoms is approximately proportional to $Z^2$ ($Z$: atomic number), only heavier La and Ni atoms are visible while lighter O atoms remain invisible. Low magnification STEM-HAADF images clearly depict the layered boundaries within these heterostructures. The LSNO$_3$ grown on the TiO$_2$-terminated surface exhibits a standard perovskite structure without any detectable atomic dislocations (Fig. 2A). In contrast, atomic disarrangements are observed at both LSNO$_3$/STO(mixed) and LSNO$_3$/SAO$_T$ interfaces due to ion diffusion and surface defects as depicted in Fig. 2 (B and C). Within these regions indicated by dark red rectangles, the atomic contrast between *A* sites and *B* sites of the *AB*O$_3$ perovskite structure is comparable, suggesting the possible presence of stacking faults, primarily of the Ruddlesden-Popper-type (RP-type) faults (*3, 44*). Other regions labeled with orange rectangles appear disordered with low contrast, implying an absence of ordered structure formation. The STEM-HAADF experiments demonstrate the crystalline quality of LSNO$_3$ layer grown on the TiO$_2$-terminated surface, thereby confirming the effectiveness of the atomically interface engineering from a microscopic perspective.

The three types of heterostructure samples, together with CaH$_2$ tablets, were loaded in a home-made vacuum chamber and annealed under the same conditions (More details shown in the Materials and Methods Section). Following topotactic reduction, the apical oxygen atoms in the octahedral structure were removed, leading to the formation of the well-confined infinite-layer structure La$_{0.8}$Sr$_{0.2}$NiO$_2$ (LSNO$_2$) from the perovskite phase (Fig. 3). The XRD analysis of the LSNO$_2$/STO(TiO$_2$)/SAO$_T$ heterostructure reveals distinct LSNO$_2$ diffraction peaks at (001) and (002), with a comparable out-of-plane $c$ lattice constant to that of superconducting nickelate films on substrates. In contrast, due to the presence of RP-type faults and other crystalline defects, oxygen atoms were randomly removed and no clear infinite-layer structures were observed in the other two types of heterostructures, which is supported by the absence of the infinite-layer structure $LSNO_2$ diffraction peaks in XRD $2\theta$–$\omega$ scans (Fig. 3B). Note that the structure of the $SAO_T$ sacrificial layer is well preserved, ensuring its water-soluble property through the topotactic reduction process.

The $LSNO_2/STO(TiO_2)/SAO_T$ samples were attached onto Polydimethylsiloxane (PDMS) and subsequently transferred onto silicon wafer to obtain the $LSNO_2/STO(TiO_2)$ freestanding membranes after dissolving water-soluble sacrificial layer. Due to the presence of more discrete Al-O networks and a higher concentration of water-soluble Sr-O within, $SAO_T$ was found to exhibit a dissolution rate approximately 10 times faster than that of the commonly-used $Sr_3Al_2O_6$ ($SAO_C$) (*49, 50*). With the aid of $SAO_T$ sacrificial layer, the transfer process can be completed within 2 minutes, minimizing the possibility of oxygen backfilling in the freestanding membrane. In the freestanding membranes, the out-of-plane $c$ lattice constant decreases from 3.428 Å (before transfer) to 3.420 Å while the in-plane $a/b$ lattice constant increases from 3.905 Å (before transfer) to 3.917 Å/3.916 Å, as extracted through analysis of the triangular relationship of $LSNO_2$ (002) and $LSNO_2$ (101) reflections (Fig. 4, B and C). The changes in lattice constants could be attributed to the elimination of the clamping effect with an in-plane compressive epitaxial strain of 1.34% induced by STO substrate ($a_{STO\text{-bulk}}$ = 3.905 Å, $a_{LNO_2\text{-bulk}}$ = 3.958 Å (*45*)). As the $STO(TiO_2)$ buffer layer in the freestanding membrane shares the same in-plane lattice constants with the $LSNO_2$ layer (Fig. 4D), the balance of the elastic energies of the two layers results in a partially relaxed state of $LSNO_2$. To check the impact of the exposure time to water on the quality of the infinite-layer nickelate membranes, we also intentionally leave the sample in water for about 30 minutes. With extending dissolution time, the $LSNO_2$ diffraction peak position of the freestanding membrane exhibits an obvious shift towards lower $2\theta$ value in the $2\theta$–$\omega$ scans along the (00*l*) direction, reflecting an increased out-of-plane $c$ lattice constant (fig. S3) with an insulating state (not shown). It is most likely that the oxygen backfilling occurred in the freestanding membrane, resulting in the formation of $LSNO_{2+\delta}$ phase, accompanied by the expansion of crystal lattices. The striking difference of the sample quality shows the importance in minimizing the amount of exposure time to water, which favors the sacrificial layer $SAO_T$ with fast dissolution time.

Due to the micro size of the freestanding membranes, microfabrication techniques are required to prepare metal electrodes for the transport measurements. However, the typical pre-bake and post-exposure-bake steps during photolithography process would damage the metastable infinite-layer structure in the freestanding membranes. Additionally, vacuum evaporation of metals may cause crystal lattice damage at the interface due to cluster bombardment and intense local heating (*51-53*). As such, insulating behavior in freestanding membranes was observed through traditional metal electrode preparation methods (not shown). To avoid such sample degradations, an electrode transfer technique (*54-56*) was employed to preserve superconductivity in freestanding membranes. In this scheme, Au films were first deposited onto the $SiO_2/Si$ wafer using electron-beam evaporation, with electrode patterns defined by photolithography. Subsequently, the polyvinyl alcohol (PVA) film was coated onto the surface of Au electrodes. Taking the advantage of the weak van der Waals force at the $Au/SiO_2$ interface, Au electrodes can be transferred with nearly 100% yield onto the PVA film when it was peeled off from the $SiO_2/Si$ substrate. The PVA/Au hybrid was finally printed onto the target $LSNO_2$ freestanding membrane. Dissolved by deionized water within 30 seconds without local heating, Au electrodes can be transferred on the freestanding membrane for transport measurements.

The electrical transport properties of the $LSNO_2/STO(TiO_2)$ freestanding membrane on $SiO_2/Si$ wafer were characterized through standard four-terminal resistance measurements (Fig. 5). The *R-T* curve exhibits metallic behavior with comparable resistance values to the pristine sample above superconducting transition temperature. A superconducting transition begins at an onset temperature $T_c^{onset}$ of 12.5 K and hits at a zero-resistance temperature $T_c^{zero}$ of 10.9 K, where $T_c^{onset}$ is defined as the intersection of two linear extrapolations from the normal state and the transition. In contrast, the pristine film on the substrate shows slightly higher $T_c^{onset}$ = 16.6 K and $T_c^{zero}$=11.4 K. The reduced $T_c$ in freestanding membranes could be attributed to the strain relaxation (*40, 57-60*). Furthermore, we extract the upper critical field $H_{c2}$ using a criterion of 50% $R_n$ (*T*), which is determined by linear fitting to the normal state resistance between 25 K and 30 K. Evidently, superconductivity exhibits clear anisotropic behavior for different magnetic-field orientations, consistent with previous report (*61*). According to calculations based on the linearized Ginzburg–Landau (G-L) model (*62*), we determine the in-plane G-L coherence length at zero temperature $\xi_{ab}(0)$ of 32.04 Å (fig. S4), which is comparable to that observed in films grown on substrates (*61, 63*).

## Discussion

Previous works have demonstrated the challenging of the precursor phase growth and the subsequent topotactic reduction (*6, 10, 43, 64*). To balance the lattice mismatch between the substrates and the perovskite phase/the reduced infinite-layer phase, lanthanum nickelates were carefully chosen in our work. The substrate-like condition was created for the precursor phase growth via introducing a $TiO_2$-terminated $SrTiO_3$ buffer layer on the sacrificial layer. Together with the insertion of an extra [LaO] layer at the polar/non-polar ($LSNO_3$/STO) interface, the typical Ruddlesden–Popper-type faults are absent in the precursor phase nickelates $La_{0.8}Sr_{0.2}NiO_3$ films. A quicker dissolution process facilitated by $Sr_4Al_2O_7$ sacrificial layer and the well-controlled electrode transfer technique are both crucial for maintaining the metastable infinite-layer structure. By implementing these improvements, we have successfully achieved superconducting freestanding nickelate membranes. This method could be also extended to the Ruddlesden–Popper nickelate system and other materials with demanding growth requirement as well.

With more degrees of freedom, the superconducting freestanding membrane can be attached to various external platforms, enabling the application of continuous uniaxial/biaxial strain (*39, 65*) or isostatic pressure (*16*). Further investigations on the strain/pressure engineering may yield higher $T_c$ values for both infinite-layer structure nickelate and Ruddlesden–Popper nickelate systems (*17*), thereby advancing our comprehension of the underlying physics and facilitating potential applications. Moreover, the stacking and assembling properties of superconducting freestanding membranes provide a promising avenue for the fabrication of phase-sensitive devices (*18, 19, 66*). The construction of Josephson junction and angle twisted membranes is aimed at exploring the pairing symmetry, as well as the topological superconducting state with charge-4$e$ (*67, 68*) in nickelates. Nonetheless, special protection and treatments are required to preserve the surface quality of these membranes to construct junctions with clean and sharp interface. In short, the synthesis of high-quality metastable oxide membranes offers new possibilities for exploring unconventional superconductivity in nickelates and novel quantum phases in correlated oxides.

## Materials and Methods

### *Film growth and Sample Preparation:*
All the samples were grown by a DCA R450 MBE system. Before growth, commercial $SrTiO_3$ (STO) (001) substrates (10 × 10 × 1 $mm^3$) were etched in buffered 8% HF acid for 70 seconds and then annealed in oxygen at 1000 °C for 80 minutes to obtain a $TiO_2$-terminated step-and-terrace surface for film growth. The 3-unit-cell $Sr_4Al_2O_7$ ($SAO_T$) layers were epitaxially deposited under an oxygen partial pressure of 1 × $10^{-6}$ Torr at a substrate temperature of 850 °C. The STO buffer layer with $TiO_2$-termination and the 20-unit-cell $La_{0.8}Sr_{0.2}NiO_3$ ($LSNO_3$) layer were epitaxially deposited under an oxidant background pressure of ~1 ×10$^{-5}$ Torr using the distilled ozone at a substrate temperature of 600 °C. The Sr doping level was set at 20% to ensure the highest superconducting transition temperature (*15*). The nominal beam flux ratio between La and Ni was roughly measured using the quartz crystal microbalance and precisely optimized to meet the stoichiometry.

To obtain the infinite-layer phase La$_{0.8}$Sr$_{0.2}$NiO$_2$ (LSNO$_2$), the precursor phase films together with CaH$_2$ tablets (0.1 g) were loaded in a home-made vacuum chamber with background pressure lower than 1×10$^{-3}$ Torr, and then heated to 300 °C for 4 hours, with warming/cooling rate of 15 °C min$^{-1}$.

*Structural Characterizations*:

Reflection high-energy electron diffraction (RHEED) with 15 kV acceleration voltage was employed to precisely monitor the thickness as well as the crystalline quality of the as-grown films. The structure diffraction peaks were characterized by high-resolution X-ray diffraction (XRD) using a Bruker D8 Discover diffractometer, with the Cu-K$_\alpha$ radiation of 1.5418 Å wavelength.

*STEM experiments*:

The cross-sectional transmission electron microscopy (TEM) lamellae were prepared via focused ion beam using a FEI HELIOS G4 CX dual-beam focused ion beam device. Carbon and platinum protective layers were deposited onto the surface of the film during the preparation procedure. The lamellae were further polished with a voltage of 2 kV and a beam current of 9 pA. The scanning transmission electron microscopy in high-angle annular dark field mode (STEM-HAADF) images were obtained using an aberration corrected FEI Titan Cubed G2 60-300 with an accelerating voltage of 300 kV and a convergence semi-angle of 22.5 mrad, in conjunction with a JEOL ARM 200F with a 200 kV Gun and a 27.2 mrad convergence semi-angle.

*Transfer Method*:

First, the top surface of the LSNO$_2$ film grown on SAO$_T$ sacrificial layer was attached to a flexible polymer substrate - polydimethylsiloxane (PDMS); Second, the sample was immersed in de-ionized water for two minutes to dissolve the water-soluble SAO$_T$ sacrificial layer; Third, the freestanding LSNO$_3$/STO(TiO$_2$) membrane on PDMS was dried and attached to a silicon wafer with secondary transfer; Finally, the freestanding film remained on the silicon wafer after peeling off the PDMS.

*Transport Measurements*:

Transport measurements down to 2 K were performed at the integrated Cryofree® superconducting magnet system equipped with a 12 T magnetic field, using standard a.c. lock-in techniques at ~13.3 Hz. The ohmic contacts were made by the prepared electrodes and the ultrasonic aluminum-wire bonding. The magnetic field were applied by aligning along and perpendicular to the *c* axis of the film.

**References**

1. D. Li, K. Lee, B.Y. Wang, M. Osada, S. Crossley, H.R. Lee, Y. Cui, Y. Hikita, H.Y. Hwang, Superconductivity in an infinite-layer nickelate. *Nature.* **572,** 624-627 (2019).

2. B. Keimer, S.A. Kivelson, M.R. Norman, S. Uchida, J. Zaanen, From quantum matter to high-temperature superconductivity in copper oxides. *Nature.* **518,** 179-186 (2015).

3. C.C. Tam, J. Choi, X. Ding, S. Agrestini, A. Nag, M. Wu, B. Huang, H. Luo, P. Gao, M. Garcia-Fernandez, L. Qiao, K.J. Zhou, Charge density waves in infinite-layer NdNiO$_2$ nickelates. *Nat. Mater.* **21,** 1116–1120 (2022).

4. G. Krieger, L. Martinelli, S. Zeng, L.E. Chow, K. Kummer, R. Arpaia, M. Moretti Sala, N.B. Brookes, A. Ariando, N. Viart, M. Salluzzo, G. Ghiringhelli, D. Preziosi, Charge and Spin Order Dichotomy in NdNiO$_2$ Driven by the Capping Layer. *Phys. Rev. Lett.* **129,** 027002 (2022).

5. M. Rossi, M. Osada, J. Choi, S. Agrestini, D. Jost, Y. Lee, H. Lu, B.Y. Wang, K. Lee, A. Nag, Y.-D. Chuang, C.-T. Kuo, S.-J. Lee, B. Moritz, T.P. Devereaux, Z.-X. Shen, J.-S. Lee, K.-J.

Zhou, H.Y. Hwang, W.-S. Lee, A broken translational symmetry state in an infinite-layer nickelate. *Nat. Phys.* **18,** 869-873 (2022).

6. K. Lee, B.Y. Wang, M. Osada, B.H. Goodge, T.C. Wang, Y. Lee, S. Harvey, W.J. Kim, Y. Yu, C. Murthy, S. Raghu, L.F. Kourkoutis, H.Y. Hwang, Linear-in-temperature resistivity for optimally superconducting (Nd,Sr)NiO$_2$. *Nature.* **619,** 288-292 (2023).

7. J. Fowlie, M. Hadjimichael, M.M. Martins, D. Li, M. Osada, B.Y. Wang, K. Lee, Y. Lee, Z. Salman, T. Prokscha, J.-M. Triscone, H.Y. Hwang, A. Suter, Intrinsic magnetism in superconducting infinite-layer nickelates. *Nat. Phys.* **18,** 1043-1047 (2022).

8. H. Lu, M. Rossi, A. Nag, M. Osada, D.F. Li, K. Lee, B. Y. Wang, M. Garcia-Fernandez, S. Agrestini, Z. X. Shen, E. M. Been, B. Moritz, T. P. Devereaux, J. Zaanen, H. Y. Hwang, Ke-Jin Zhou, W.S. Lee, Magnetic excitations in infinite-layer nickelates. *Science.* **373,** 213-216 (2021).

9. J.G.Bednorz, K.A. Miiller, Possible High $T_c$ Superconductivity in the Ba - La- Cu- O System. *Z. Phys. B.* **64,** 189-193 (1986).

10. D. Li, B.Y. Wang, K. Lee, S.P. Harvey, M. Osada, B.H. Goodge, L.F. Kourkoutis, H.Y. Hwang, Superconducting Dome in Nd$_{1-x}$Sr$_x$NiO$_2$ Infinite Layer Films. *Phys. Rev. Lett.* **125,** 027001 (2020).

11. S. Zeng, C.S. Tang, X. Yin, C. Li, M. Li, Z. Huang, J. Hu, W. Liu, G.J. Omar, H. Jani, Z.S. Lim, K. Han, D. Wan, P. Yang, S.J. Pennycook, A.T.S. Wee, A. Ariando, Phase Diagram and Superconducting Dome of Infinite-Layer Nd$_{1-x}$Sr$_x$NiO$_2$ Thin Films. *Phys. Rev. Lett.* **125,** 147003 (2020).

12. M. Osada, B.Y. Wang, B.H. Goodge, K. Lee, H. Yoon, K. Sakuma, D. Li, M. Miura, L.F. Kourkoutis, H.Y. Hwang, A Superconducting Praseodymium Nickelate with Infinite Layer Structure. *Nano Lett.* **20,** 5735-5740 (2020).

13. M. Osada, B.Y. Wang, K. Lee, D. Li, H.Y. Hwang, Phase diagram of infinite layer praseodymium nickelate Pr$_{1-x}$Sr$_x$NiO$_2$ thin films. *Phys. Rev. Mater.* **4,** 121801 (2020).

14. Shengwei Zeng, Changjian Li, Lin Er Chow, Z.Z. Yu Cao, Chi Sin Tang, Xinmao Yin, Zhi Shiuh Lim, Junxiong Hu, Ping Yang, Ariando Ariando, Superconductivity in infinite-layer nickelate La$_{1-x}$Ca$_x$NiO$_2$ thin films. *Sci. Adv,.* **8,** eabl9927 (2022).

15. M. Osada, B.Y. Wang, B.H. Goodge, S.P. Harvey, K. Lee, D. Li, L.F. Kourkoutis, H.Y. Hwang, Nickelate Superconductivity without Rare-Earth Magnetism: (La,Sr)NiO$_2$. *Adv. Mater.* **33,** e2104083 (2021).

16. N.N. Wang, M.W. Yang, Z. Yang, K.Y. Chen, H. Zhang, Q.H. Zhang, Z.H. Zhu, Y. Uwatoko, L. Gu, X.L. Dong, J.P. Sun, K.J. Jin, J.G. Cheng, Pressure-induced monotonic enhancement of $T_c$ to over 30 K in superconducting Pr$_{0.82}$Sr$_{0.18}$NiO$_2$ thin films. *Nat. Commun.* **13,** 4367 (2022).

17. H. Sun, M. Huo, X. Hu, J. Li, Z. Liu, Y. Han, L. Tang, Z. Mao, P. Yang, B. Wang, J. Cheng, D.-X. Yao, G.-M. Zhang, M. Wang, Signatures of superconductivity near 80 K in a nickelate under high pressure. *Nature.* **621,** 493-498 (2023).

18. Q. Gu, Y. Li, S. Wan, H. Li, W. Guo, H. Yang, Q. Li, X. Zhu, X. Pan, Y. Nie, H.H. Wen, Single particle tunneling spectrum of superconducting Nd$_{1-x}$Sr$_x$NiO$_2$ thin films. *Nat. Commun.* **11,** 6027 (2020).

19. L. E. Chow, S. Kunniniyil Sudheesh, Z. Y. Luo, P. Nandi, T. Heil, J. Deuschle, S. W. Zeng, Z. T. Zhang, S. Prakash, X. M. Du, Z. S. Lim, Peter A. van Aken, Elbert E. M. Chia, A. Ariando, Pairing symmetry in infinite-layer nickelate superconductors. *arxiv.* **2201.10038** (2023).

20. B. Cheng, D. Cheng, K. Lee, L. Luo, Z. Chen, Y. Lee, B.Y. Wang, M. Mootz, I.E. Perakis, Z.-X. Shen, H.Y. Hwang, J. Wang, Evidence for *d*-wave superconductivity of infinite-layer nickelates from low-energy electrodynamics. *Nat. Mater.* (2024).

21. F. Lechermann, Multiorbital Processes Rule the $Nd_{1-x}Sr_xNiO_2$ Normal State. *Phys. Rev. X.* **10,** 041002 (2020).

22. I. Leonov, S.L. Skornyakov, S.Y. Savrasov, Lifshitz transition and frustration of magnetic moments in infinite-layer $NdNiO_2$ upon hole doping. *Phys. Rev. B.* **101,** 241108 (2020).

23. H. Sakakibara, H. Usui, K. Suzuki, T. Kotani, H. Aoki, K. Kuroki, Model Construction and a Possibility of Cuprate like Pairing in a New $d^9$ Nickelate Superconductor $(Nd,Sr)NiO_2$. *Phys. Rev. Lett.* **125,** 077003 (2020).

24. X. Wu, D. Di Sante, T. Schwemmer, W. Hanke, H.Y. Hwang, S. Raghu, R. Thomale, Robust $d_{x2-y2}$-wave superconductivity of infinite-layer nickelates. *Phys. Rev. B.* **101,** 060504 (2020).

25. M. Kitatani, L. Si, O. Janson, R. Arita, Z. Zhong, K. Held, Nickelate superconductors—a renaissance of the one-band Hubbard model. *npj Quantum Mater.* **5,** 59 (2020).

26. Q. Li, C. He, J. Si, X. Zhu, Y. Zhang, H.-H. Wen, Absence of superconductivity in bulk $Nd_{1-x}Sr_xNiO_2$. *Commun. Mater.* **1,** 16 (2020).

27. B.-X. Wang, H. Zheng, E. Krivyakina, O. Chmaissem, P.P. Lopes, J.W. Lynn, L.C. Gallington, Y. Ren, S. Rosenkranz, J.F. Mitchell, D. Phelan, Synthesis and characterization of bulk $Nd_{1-x}Sr_xNiO_2$ and $Nd_{1-x}Sr_xNiO_3$. *Phys. Rev. Mater.* **4,** 084409 (2020).

28. M. Huo, Z. Liu, H. Sun, L. Li, H. Lui, C. Huang, F. Liang, B. Shen, M. Wang, Synthesis and properties of $La_{1-x}Sr_xNiO_3$ and $La_{1-x}Sr_xNiO_2$. *Chinese Phys B.* **31,** 107401 (2022).

29. D.A. Wollman, D.J. Van Harlingen, W.C. Lee, D.M. Ginsberg, A.J. Leggett, Experimental determination of the superconducting pairing state in YBCO from the phase coherence of YBCO-Pb dc SQUIDs. *Phys. Rev. Lett.* **71,** 2134-2137 (1993).

30. C.C.T., J.R. Kirtley, Pairing symmetry in cuprate superconductors. *Rev. Mod. Phys.* **72,** 969 (2000).

31. X. Xu, Y. Li, C.L. Chien, Spin-Triplet Pairing State Evidenced by Half-Quantum Flux in a Noncentrosymmetric Superconductor. *Phys. Rev. Lett.* **124,** 167001 (2020).

32. Yufan Li, Xiaoying Xu, M.-H. Lee, M.-W. Chu, C.L. Chien, Observation of half-quantum flux in the unconventional superconductor $\beta$-$Bi_2Pd$. *Science.* **366,** 238-241 (2019).

33. D. Lu, D.J. Baek, S.S. Hong, L.F. Kourkoutis, Y. Hikita, Harold Y. Hwang, Synthesis of freestanding single-crystal perovskite films and heterostructures by etching of sacrificial water-soluble layers. *Nat. Mater.* **15,** 1255-1260 (2016).

34. D. Ji, S. Cai, T.R. Paudel, H. Sun, C. Zhang, L. Han, Y. Wei, Y. Zang, M. Gu, Y. Zhang, W. Gao, H. Huyan, W. Guo, D. Wu, Z. Gu, E.Y. Tsymbal, P. Wang, Y. Nie, X. Pan, Freestanding crystalline oxide perovskites down to the monolayer limit. *Nature.* **570,** 87-90 (2019).

35. H. Sun, J. Wang, Y. Wang, C. Guo, J. Gu, W. Mao, J. Yang, Y. Liu, T. Zhang, T. Gao, H. Fu, T. Zhang, Y. Hao, Z. Gu, P. Wang, H. Huang, Y. Nie, Nonvolatile ferroelectric domain wall memory integrated on silicon. *Nat. Commun.* **13,** 4332 (2022).

36. Guohua Dong, Suzhi Li, Mouteng Yao, Ziyao Zhou, Yong-Qiang Zhang, Z.L. Xu Han, Junxiang Yao, Bin Peng, Zhongqiang Hu, Houbing Huang, Tingting Jia, Jiangyu Li, Wei Ren, Zuo-Guang Ye, Xiangdong Ding, Jun Sun, Ce-Wen Nan, Long-Qing Chen, Ju Li, Ming Liu, Super-elastic ferroelectric single-crystal membrane with continuous electric dipole rotation. *Science.* **366,** 475-479 (2020).

37. Z.L. Wang, J. Song, Piezoelectric Nanogenerators Basedon Zinc Oxide Nanowire Arrays. *Science.* **312,** 242-246 (2006).

38. Y. Zhu, M. Liao, Q. Zhang, H.-Y. Xie, F. Meng, Y. Liu, Z. Bai, S. Ji, J. Zhang, K. Jiang, R. Zhong, J. Schneeloch, G. Gu, L. Gu, X. Ma, D. Zhang, Q.-K. Xue, Presence of *s*-Wave Pairing in Josephson Junctions Made of Twisted Ultrathin $Bi_2Sr_2CaCu_2O_{8+x}$ Flakes. *Phys. Rev. X.* **11,** 031011 (2021).

39. Seung Sae Hong, Mingqiang Gu, ManishVerma, Varun Harbola, Bai Yang Wang, DiLu, Arturas Vailionis, Yasuyuki Hikita, Rossitza Pentcheva, James M. Rondinelli, Harold Y. Hwang, Extreme tensile strain states in $La_{0.7}Ca_{0.3}MnO_3$ membranes. *Science.* **368,** 71-76 (2020).

40. P. Malinowski, Q. Jiang, J.J. Sanchez, J. Mutch, Z. Liu, P. Went, J. Liu, P.J. Ryan, J.-W. Kim, J.-H. Chu, Suppression of superconductivity by anisotropic strain near a nematic quantum critical point. *Nat. Phys.* **16,** 1189-1193 (2020).

41. Clifford W. Hicks, Daniel O. Brodsky, Edward A. Yelland, Alexandra S. Gibbs, JanA.N.Bruin, Mark E. Barber, Stephen D. Edkins, Keigo Nishimura, Shingo Yonezawa, Yoshiteru Maeno, Andrew P. Mackenzie, Strong Increase of $T_c$ of $Sr_2RuO_4$ Under Both Tensile and Compressive Strain. *Science.* **344,** 283-285 (2014).

42. C. Adamo, X. Ke, H.Q. Wang, H.L. Xin, T. Heeg, M.E. Hawley, W. Zander, J. Schubert, P. Schiffer, D.A. Muller, L. Maritato, D.G. Schlom, Effect of biaxial strain on the electrical and magnetic properties of (001) $La_{0.7}Sr_{0.3}MnO_3$ thin films. *Appl. Phys. Lett.* **95,** 112504 (2009).

43. K. Lee, B.H. Goodge, D. Li, M. Osada, B.Y. Wang, Y. Cui, L.F. Kourkoutis, H.Y. Hwang, Aspects of the synthesis of thin film superconducting infinite-layer nickelates. *APL Mater.* **8,** 041107 (2020).

44. Y. Li, X. Cai, W. Sun, J. Yang, W. Guo, Z. Gu, Y. Zhu, Y. Nie, Synthesis of Chemically Sharp Interface in $NdNiO_3/SrTiO_3$ Heterostructures. *Chinese Phys. Lett.* **40,** 076801 (2023).

45. M. A. Hayward, M. A. Green, M. J. Rosseinsky, J.S. Sodium, Sodium Hydride as a Powerful Reducing Agent for Topotactic Oxide Deintercalation: Synthesis and Characterization of the Nickel(I) Oxide $LaNiO_2$. *J. Am. Chem. Soc.* **121,** 8843-8854 (1999).

46. M. Kawai, S. Inoue, M. Mizumaki, N. Kawamura, N. Ichikawa, Y. Shimakawa, Reversible changes of epitaxial thin films from perovskite $LaNiO_3$ to infinite-layer structure $LaNiO_2$. *Appl. Phys. Lett.* **94,** 082102 (2009).

47. H.Y. Sun, Z.W. Mao, T.W. Zhang, L. Han, T.T. Zhang, X.B. Cai, X. Guo, Y.F. Li, Y.P. Zang, W. Guo, J.H. Song, D.X. Ji, C.Y. Gu, C. Tang, Z.B. Gu, N. Wang, Y. Zhu, D.G. Schlom, Y.F. Nie, X.Q. Pan, Chemically specific termination control of oxide interfaces via layer-by-layer mean inner potential engineering. *Nat. Commun.* **9,** 2965 (2018).

48. Y. Li, W. Sun, J. Yang, X. Cai, W. Guo, Z. Gu, Y. Zhu, Y. Nie, Impact of Cation Stoichiometry on the Crystalline Structure and Superconductivity in Nickelates. *Front. Phys.* **9,** 719534 (2021).

49. Jinfeng Zhang, Ting Lin, Ao Wang, Xiaochao Wang, Qingyu He, Huan Ye, Jingdi Lu, Qing Wang, Zhengguo Liang, Feng Jin, Shengru Chen, Minghui Fan, Er-Jia Guo, Qinghua Zhang, Lin Gu, Zhenlin Luo, Liang Si, Wenbin Wu, Lingfei Wang, Super-tetragonal $Sr_4Al_2O_7$: a versatile sacrificial layer for high-integrity freestanding oxide membranes. *arXiv.* **2307.14966** (2023).

50. Leyan Nian, Haoying Sun, Zhichao Wang, Duo Xu, Hao Bo, Y. Shengjun, Yueying Li, Jian Zhou, Yu Deng, Yufeng Hao, Y. Nie, $Sr_4Al_2O_7$: A New Sacrificial Layer with High Water Dissolution Rate for the Synthesis of Freestanding Oxide Membranes. *arXiv.* **2307.14584.** (2023).

51. Recep Zan, Q. M., Ramasse, Rashid Jalil, Thanasis Georgiou, Ursel Bangert, K.S. Novoselov, Control of Radiation Damage in $MoS_2$ by Graphene Encapsulation. *ACS Nano.* **7,** 10167-10174 (2013).

52. W.A. Saidi, Influence of strain and metal thickness on metal-$MoS_2$ contacts. *J. Chem. Phys.* **141,** 094707 (2014).

53. J. Kang, W. Liu, D. Sarkar, D. Jena, K. Banerjee, Computational Study of Metal Contacts to Monolayer Transition-Metal Dichalcogenide Semiconductors. *Phys. Rev. X.* **4,** 031005 (2014).

54. Y. Liu, J. Guo, E. Zhu, L. Liao, S.-J. Lee, M. Ding, I. Shakir, V. Gambin, Y. Huang, X. Duan, Approaching the Schottky–Mott limit in van der Waals metal–semiconductor junctions. *Nature.* **557,** 696-700 (2018).

55. G. Liu, Z. Tian, Z. Yang, Z. Xue, M. Zhang, X. Hu, Y. Wang, Y. Yang, P.K. Chu, Y. Mei, L. Liao, W. Hu, Z. Di, Graphene-assisted metal transfer printing for wafer-scale integration of metal electrodes and two-dimensional materials. *Nat. Electron.* **5,** 275-280 (2022).

56. X. Yang, J. Li, R. Song, B. Zhao, J. Tang, L. Kong, H. Huang, Z. Zhang, L. Liao, Y. Liu, X. Duan, X. Duan, Highly reproducible van der Waals integration of two-dimensional electronics on the wafer scale. *Nat. Nanotech.* **18,** 471-478 (2023).

57. H. Sato, M. Naito, Increase in the superconducting transition temperature by anisotropic strain effect in (001) $La_{1.85}Sr_{0.15}CuO_4$ thin films on $LaSrAlO_4$ substrates. *Physica C Supercond.* **274,** 221 (1997).

58. T.L. Meyer, L. Jiang, S. Park, T. Egami, H.N. Lee, Strain-relaxation and critical thickness of epitaxial $La_{1.85}Sr_{0.15}CuO_4$ films. *APL Mater.* **3,** 126102 (2015).

59. Z. Jia, C.S. Tang, J. Wu, C. Li, W. Xu, K. Wu, D. Zhou, P. Yang, S. Zeng, Z. Zeng, D. Zhang, A. Ariando, M.B.H. Breese, C. Cai, X. Yin, Self-passivated freestanding superconducting oxide film for flexible electronics. *Appl. Phys. Rev.* **10,** 031401 (2023).

60. J. Shiogai, A. Tsukazaki, Superconducting FeSe membrane synthesized by etching of water-soluble $Sr_3Al_2O_6$ layer. *Appl. Phys. Lett.* **122,** 052602 (2023).

61. W. Sun, Y. Li, R. Liu, J. Yang, J. Li, W. Wei, G. Jin, S. Yan, H. Sun, W. Guo, Z. Gu, Z. Zhu, Y. Sun, Z. Shi, Y. Deng, X. Wang, Y. Nie, Evidence for Anisotropic Superconductivity Beyond Pauli Limit in Infinite-Layer Lanthanum Nickelates. *Adv. Mater.* **35,** 2303400 (2023).

62. F.E. Harper, M. Tinkham, The Mixed State in Superconducting Thin Films. *Phys. Rev.* **172,** 441-450 (1968).

63. B.Y. Wang, D. Li, B.H. Goodge, K. Lee, M. Osada, S.P. Harvey, L.F. Kourkoutis, M.R. Beasley, H.Y. Hwang, Isotropic Pauli-limited superconductivity in the infinite-layer nickelate $Nd_{0.775}Sr_{0.225}NiO_2$. *Nat. Phys.* **17,** 473-477 (2021).

64. Q. Guo, S. Farokhipoor, C. Magén, F. Rivadulla, B. Noheda, Tunable resistivity exponents in the metallic phase of epitaxial nickelates. *Nat. Commun.* **11,** 1 (2020).

65. L. Han, Y. Fang, Y. Zhao, Y. Zang, Z. Gu, Y. Nie, X. Pan, Giant Uniaxial Strain Ferroelectric Domain Tuning in Freestanding $PbTiO_3$ Films. *Adv. Mater. Interfaces.* **7,** 1901604 (2020).

66. Bing Cheng, Di Cheng, Kyuho Lee, Liang Luo, Zhuoyu Chen, Yonghun Lee, Bai Yang Wang, Martin Mootz, Ilias E. Perakis, Zhi-Xun Shen, Harold Y. Hwang, J. Wang, Low-energy electrodynamics of infinite-layer nickelates: evidence for *d*–wave superconductivity in the dirty limit. *arxiv.* **2310.02586** (2023).

67. Y.-B. Liu, J. Zhou, C. Wu, F. Yang, Charge-4e superconductivity and chiral metal in 45°-twisted bilayer cuprates and related bilayers. *Nat. Commun.* **14,** 7926 (2023).

68. S. Y. Frank Zhao, Xiaomeng Cui, Pavel A. Volkov, Hyobin Yoo, Sangmin Lee, Jules A. Gardener, Austin J. Akey, Rebecca Engelke, Yuval Ronen, Ruidan Zhong, Genda Gu, Stephan Plugge, Tarun Tummuru, Miyoung Kim, Marcel Franz, Jedediah H. Pixley, Nicola Poccia, P. Kim, Time-reversal symmetry breaking superconductivity between twisted cuprate superconductors. *Science.* **382,** 1422-1427 (2023).

**Acknowledgments**

**Funding:** This work was supported by the National Key R&D Program of China (Grant Nos. 2021YFA1400400 and 2022YFA1402502); National Natural Science Foundation of China (Grant No. 11861161004) and the Fundamental Research Funds for the Central Universities (No. 0213–14380221).

**Author contributions:** Y.F.N. conceived the project and provided scientific guidance. S.J.Y., W.J.S., Y.Y.L., H.Y.S, B.H. and W.G. prepared the thin films. W.M. conducted the STEM measurements. S.J.Y. and Y.Y.L. conducted the XRD measurements. S.J.Y. and J.F.Y. conducted the annealing experiments. S.J.Y. conducted the electrode transfer and film transfer process. W.J.S. and S.J.Y. conducted the electrical measurements. S.J.Y., W.J.S., Y.Y.L., L.Y.N., H.Y.S and Y.F.N. wrote the manuscript with contributions from all authors. All authors have discussed the results and the interpretations.

**Competing interests:** Authors declare that they have no competing interests.

**Data and materials availability:** All data are available in the main text or the supplementary materials.

**Figures and Tables**

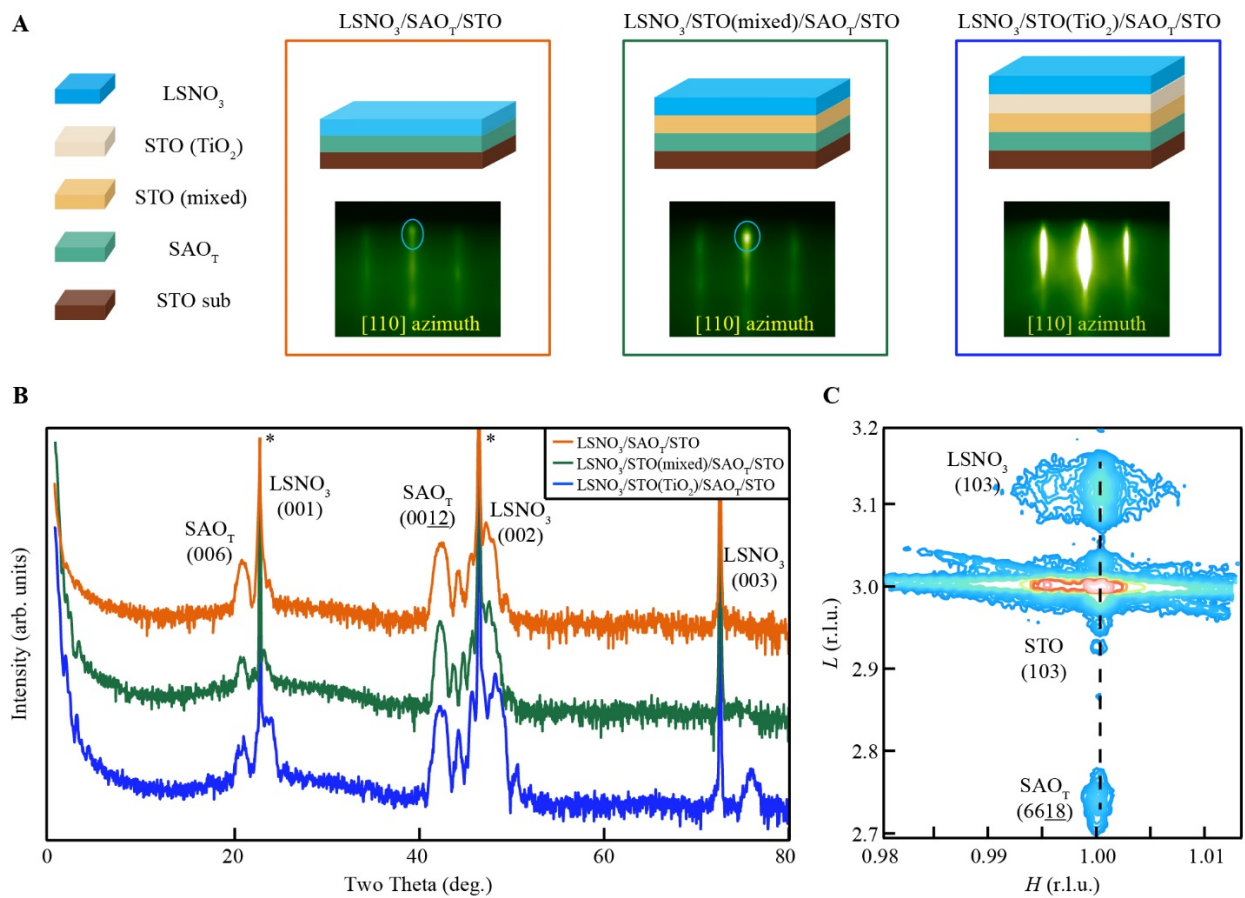

**Fig. 1. Epitaxial growth and structural characterizations of the precursor phase perovskite nickelate films on STO substrates with the insertion of the sacrificial layer.**

(**A**) Growth schemes of three types of heterostructures, namely $LSNO_3/SAO_T$, $LSNO_3/STO(mixed)/SAO_T$, $LSNO_3/STO(TiO_2)/SAO_T$ on STO substrates and their corresponding reflection high-energy electron diffraction (RHEED) patterns along [110] azimuth. The diffraction spots of the impurity phase are indicated by the blue circle. (**B**) X-ray diffraction (XRD) $2\theta$-$\omega$ scans of the heterostructures along the [00$l$] direction. (**C**) Reciprocal space mapping (RSM) images around the STO (103) reflection for the $LSNO_3/STO(TiO_2)/SAO_T$ heterostructure. Note: $LSNO_3$ represents $La_{0.8}Sr_{0.2}NiO_3$; $SAO_T$ represents $Sr_4Al_2O_7$; $STO(TiO_2)$ represents $SrTiO_3$ layer with $TiO_2$-termination; STO(mixed) represents $SrTiO_3$ layer with mixed-termination.

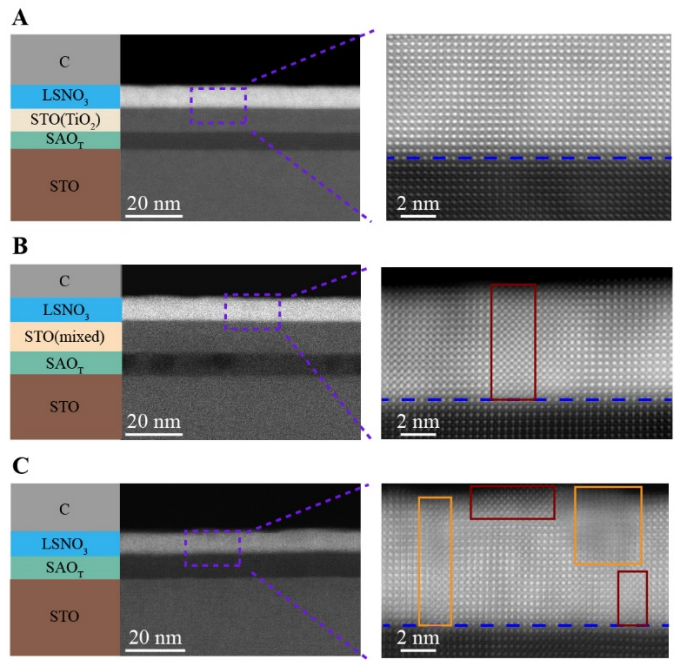

**Fig. 2. Scanning transmission electron microscopy in the high-angle annular dark field mode (STEM-HAADF) images of the heterostructures.**

(**A**, **B** and **C**) Cross-sectional STEM-HAADF images and the respective magnified views of (**A**) the LSNO$_3$/STO(TiO$_2$)/SAO$_T$ (**B**) the LSNO$_3$/STO(mixed)/SAO$_T$ (**C**) the LSNO$_3$/SAO$_T$. Defective regions are marked by rectangles and the interfaces are labelled by blue dash lines. Note: Atoms with higher and lower intensity are La and Ni, respectively.

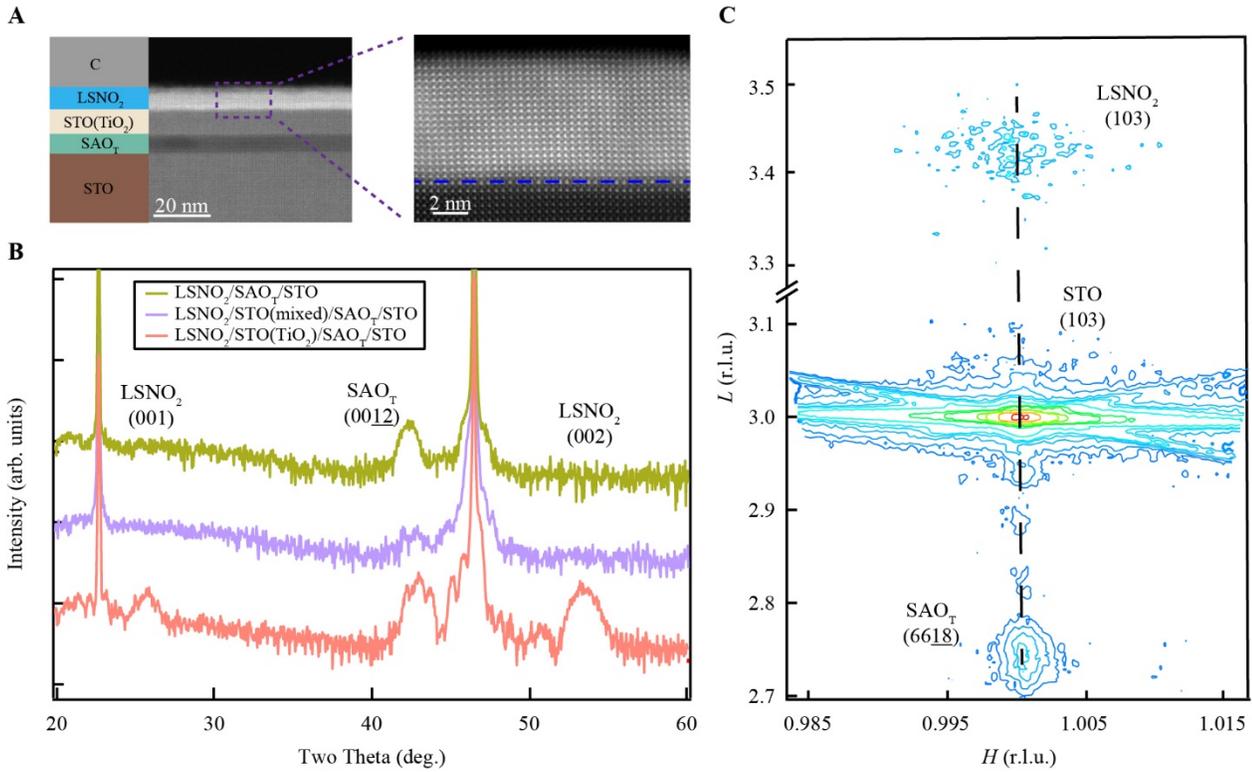

**Fig. 3. STEM-HAADF images and XRD characterizations of the infinite-layer heterostructure LSNO₂/STO(TiO₂)/SAO_T on STO substrates.**

(**A**) The cross-sectional STEM-HAADF image of the LSNO₂/STO(TiO₂)/SAO_T infinite-layer heterostructure. (**B**) XRD 2$\theta$-$\omega$ scans of the three types of heterostructures after topotactic reduction. (**C**) RSM images around the STO (103) reflection for the infinite-layer heterostructure.

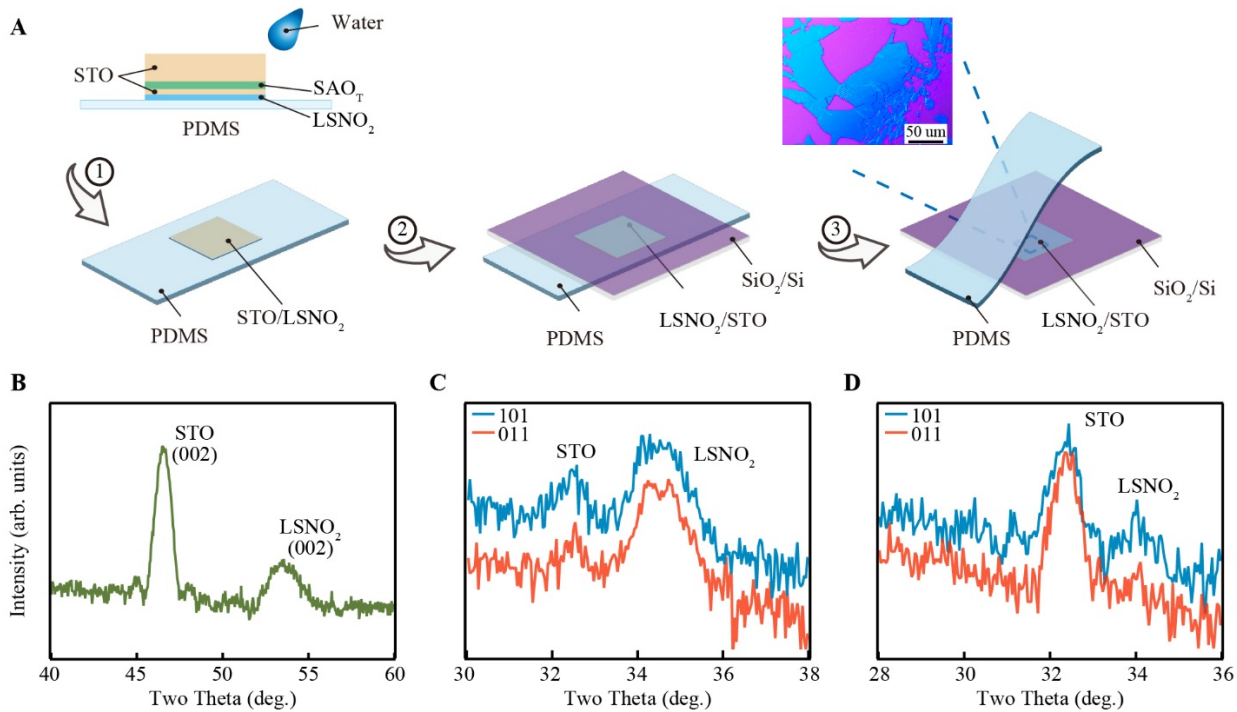

**Fig. 4. Transfer process and XRD characterizations of LSNO$_2$/STO(TiO$_2$) freestanding membranes on SiO$_2$/Si substrate.**

(**A**) Sketches of the transfer process of the freestanding membranes on SiO$_2$/Si substrate. Inset: optical image of the freestanding membrane on SiO$_2$/Si substrate. (**B**) XRD 2$\theta$-$\omega$ scans of the LSNO$_2$/STO(TiO$_2$) freestanding membrane along the [00$l$] direction. (**C** and **D**) XRD 2$\theta$-$\omega$ scans around (**C**) LSNO$_2$ (101)/(110) and (**D**) STO(101)/(110) reflections.

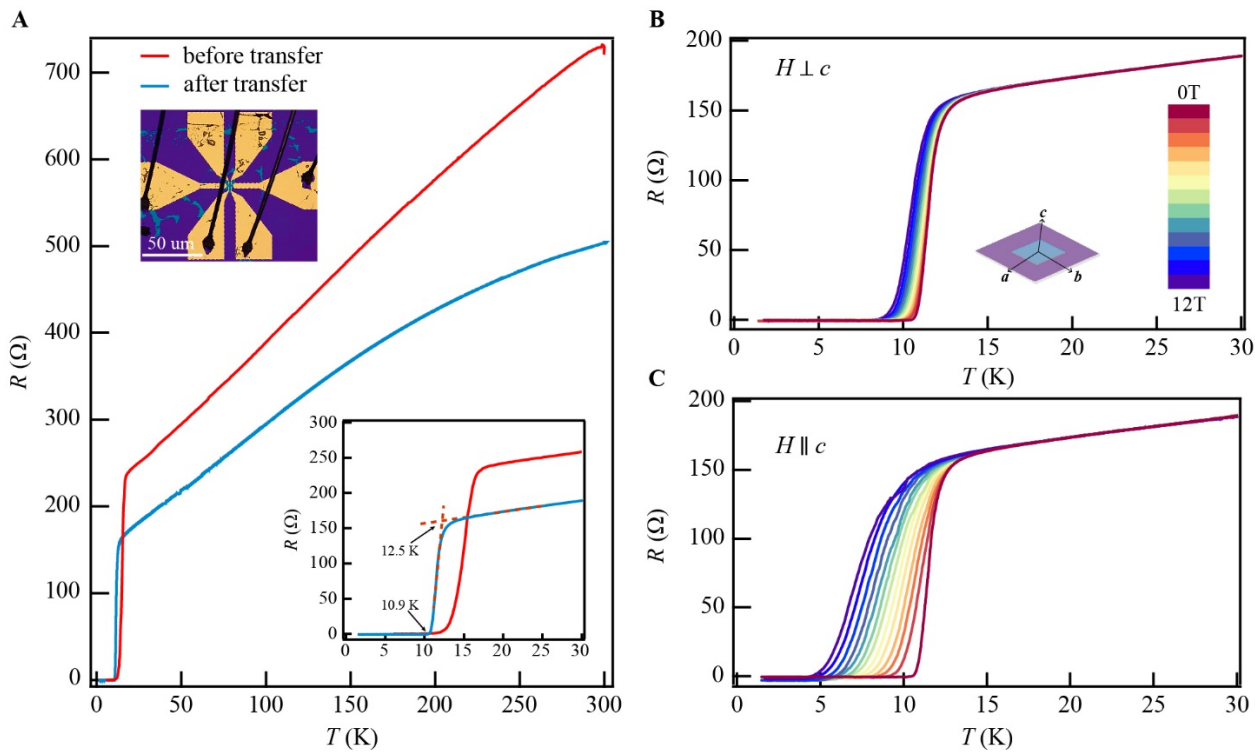

**Fig. 5. Transport measurements of the superconducting freestanding infinite-layer nickelate membranes.**

(**A**) The temperature dependence of resistance for the infinite-layer film (before transfer) and the freestanding membrane. Insets: optical image of the freestanding membrane with transferred electrodes on SiO$_2$/Si substrate and enlarged view of transport measurements near the superconducting transition temperature. (**B** and **C**) The temperature dependence of resistance under (**B**) in-plane and (**C**) out-of-plane magnetic fields up to 12 T for the freestanding membrane.

**Supplementary Materials**

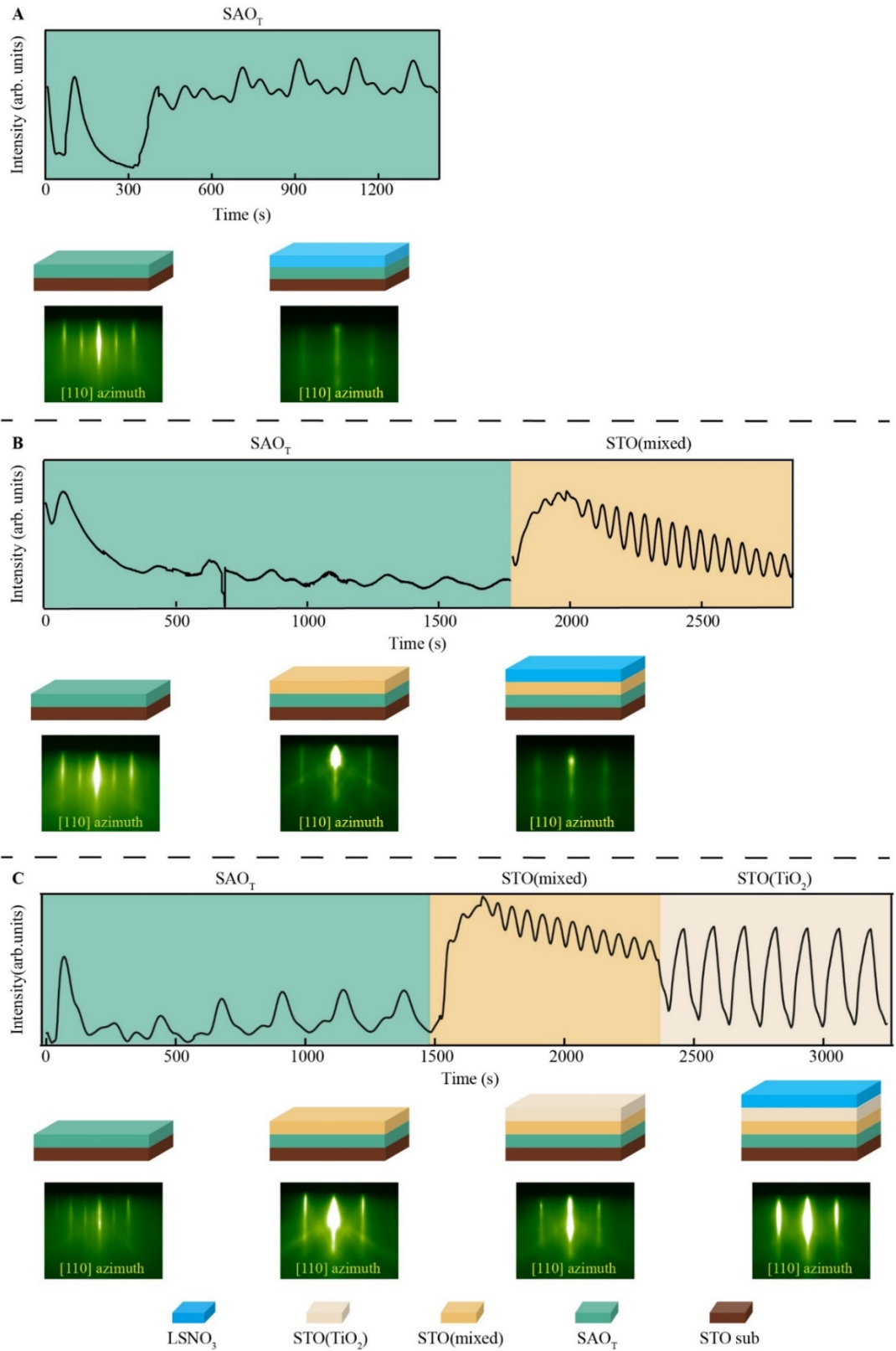

**Figure S1. Epitaxial growth schemes of the LSNO$_3$/SAO$_T$, LSNO$_3$/STO(mixed)/SAO$_T$ and LSNO$_3$/STO(TiO$_2$)/SAO$_T$ heterostructures.**

The corresponding RHEED intensity oscillations and diffraction patterns of the heterostructures (**A**) LSNO$_3$/SAO$_T$ (**B**) LSNO$_3$/STO(mixed)/SAO$_T$ (**C**) LSNO$_3$/STO(TiO$_2$)/SAO$_T$ on STO substrates.

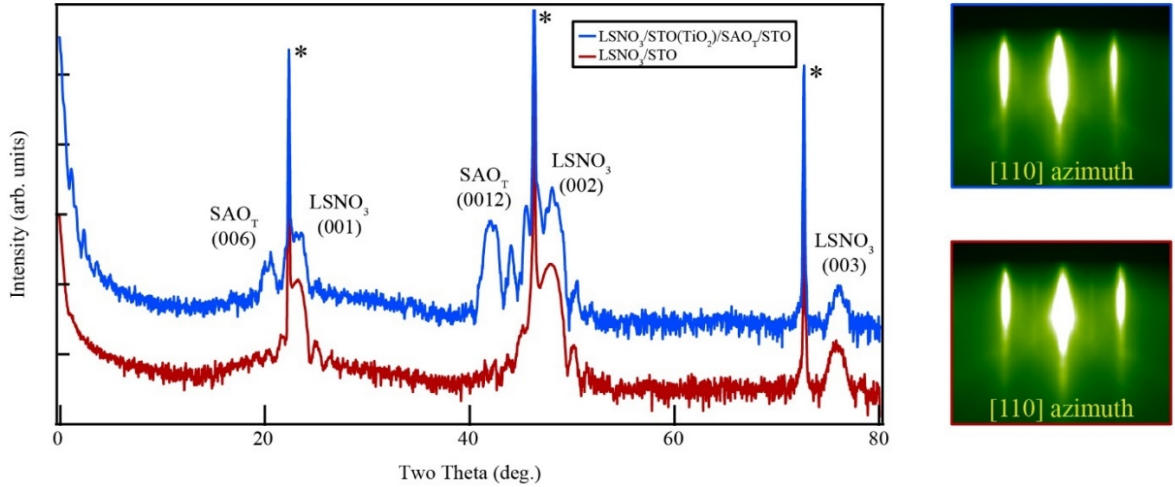

**Figure S2. XRD 2$\theta$-$\omega$ scans and the corresponding RHEED patterns of the LSNO$_3$/STO(TiO$_2$)/SAO$_T$ heterostructure and the LSNO$_3$ film grown on the STO substrate.**

The high intensity of the diffraction peaks and clear and bright RHEED patterns of the LSNO$_3$/STO(TiO$_2$)/SAO$_T$ heterostructure demonstrate the comparable crystalline quality to the nickelates directly grown on the STO substrate.

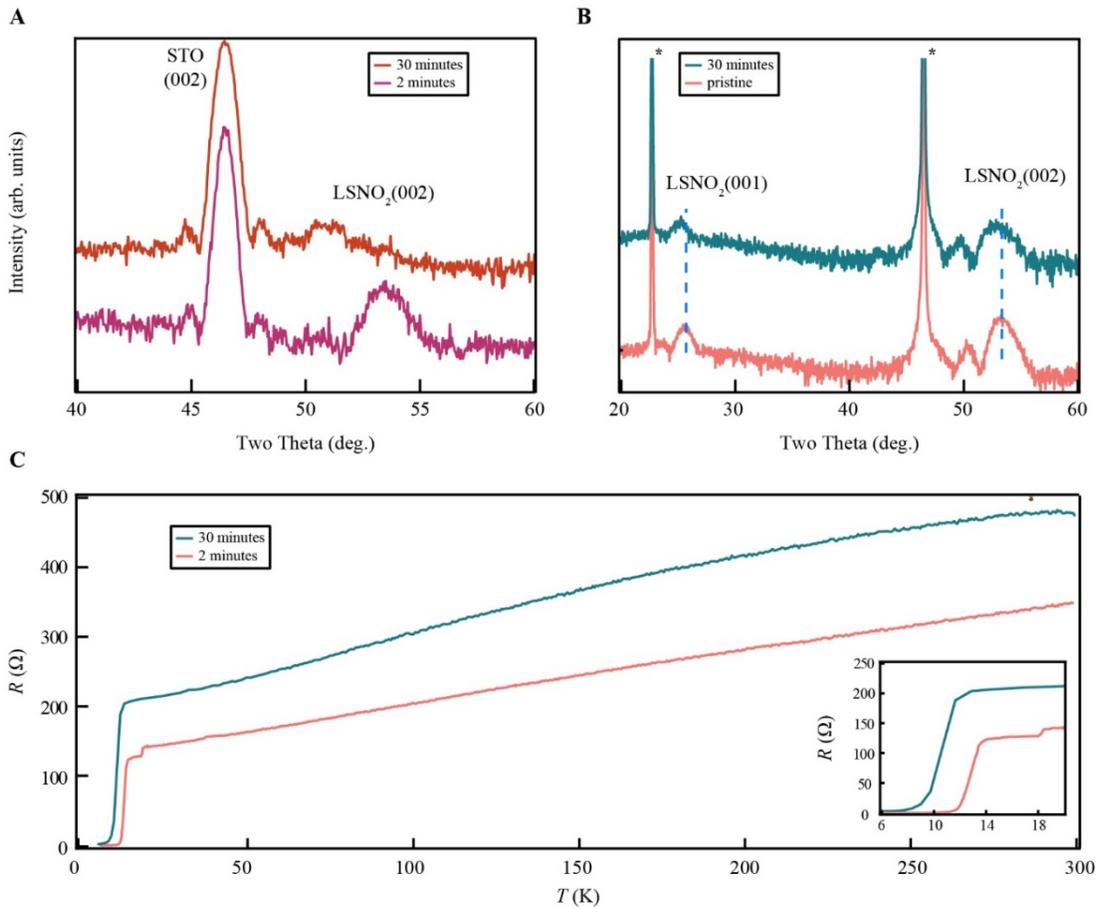

**Figure S3. Impact of dissolution time on structure and transport properties of infinite-layer structure LSNO₂ membranes and films on STO substrates.**

(**A**) XRD 2θ-ω scans of the LSNO₂ freestanding membrane with different dissolution times (30 minutes versus 2 minutes). (**B**) XRD 2θ-ω scans of the LSNO₂ films on the STO substrate with different dissolution times (30 minutes versus 2 minutes). (**C**) The temperature dependence of sheet resistance of the LSNO₂ films on the STO substrate with different dissolution times. Inset: the enlarged *R-T* curves at temperatures from 6 K to 20 K.

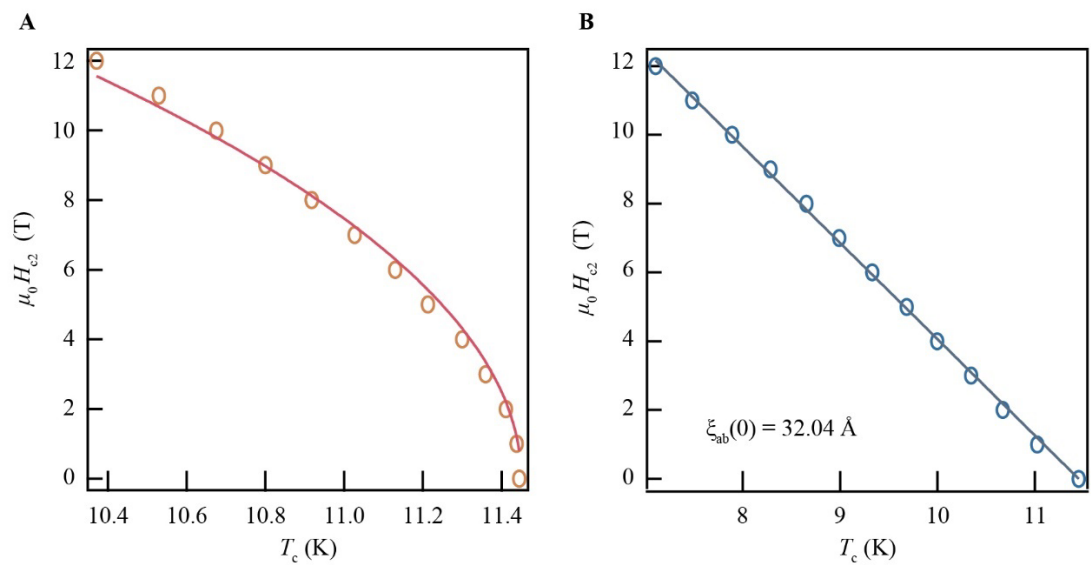

**Figure S4. Temperature-dependent upper critical field of LSNO₂ membranes.**

(**A** and **B**) Temperature-dependent upper critical field of LSNO2 membranes with magnetic fields along $c$ axis and in the $a$-$b$ plane. Here, $H_{c2}$ is defined as the field strength at which resistivity reaches 50% $R_n$ ($T$). The solid lines are corresponding fittings to $H_{c2}$ using the Ginzburg–Landau model.